\documentclass[a4paper,twoside,openany,11pt]{memoir} %openany, openright
\pdfoutput=1

\title{
    An Automated Tool for Matching Effective Theories at Finite Temperature
}
\author{
    J. Fuentes-Martín, J. López Miras, and A. Moreno-Sánchez
}
\def\fullheadfoot{0} % Use 0 for plain foots and 1 for full heads
\usepackage[english]{babel}
\usepackage{microtype,xspace}
\usepackage[utf8]{inputenc}	% Allows for writing special charachters in the tex-file 

\usepackage{amsfonts,amsmath,amssymb,bm,mathrsfs,mathtools,dsfont} 	% Standard mathematics 
\allowdisplaybreaks 	%Allows pagebreak during multiline math enviroments
\usepackage[table]{xcolor}
\usepackage{hyperref}
\usepackage{slashed}
\usepackage{multicol}

\usepackage{tikz}
\usetikzlibrary{shapes, arrows, positioning, decorations.pathmorphing, decorations.pathreplacing, decorations.shapes, decorations.text, calc} % Gauge lines
\tikzset{every picture/.style={line width=.8pt}}

\usepackage{siunitx}
\usepackage{geometry}
\usepackage[sort&compress,numbers,merge]{natbib}
\usepackage{chapterbib}
\addto\captionsenglish{\renewcommand*{\bibname}{References}}
\makeatletter
\renewcommand{\@memb@bchap}{ %\section*{\bibname} 
\bibmark \prebibhook
}
\makeatother

\usepackage{graphicx}
\graphicspath{{./Figures/}}
\usepackage{caption,subcaption}
\usepackage{multirow}

\usepackage{tabularx}
\newcolumntype{Y}{>{\centering\arraybackslash}X}
\newcolumntype{g}{>{\columncolor{black!15}}l}

\usepackage[shortlabels]{enumitem}
\setlist{itemsep=.1em,topsep=.5em}
\SetEnumerateShortLabel{i}{\textit{\roman*}}

\definecolor{red}{rgb}{0.6,.0706,.1373}
\definecolor{blue}{rgb}{0,0.396,0.741}
\definecolor{green}{rgb}{0.25,0.6,0.2}
\definecolor{teal}{rgb}{0.11,0.6,0.6}
\definecolor{orange}{rgb}{.8, .4806, 0.173}
\definecolor{yellow}{rgb}{.8, .7, 0.05}
\colorlet{blueref}{blue!80!black}
\colorlet{bluelink}{blue!90!black}
\hypersetup{
	colorlinks, 
	bookmarksopen, 
	bookmarksnumbered,
	citecolor=blueref, 		%color of links to bibliography
	linkcolor=bluelink,	%color of internal links
	urlcolor=bluelink,			%color of external links 
	pdftitle={\thetitle},    %   - title (PDF meta)
	pdfauthor={\theauthor},    	%   - author (PDF meta)
}

\addto\captionsenglish{}

\renewcommand{\contentsname}{Contents}
\renewcommand{\printtoctitle}[1]{}

\settocdepth{subsection}
\newcommand{\toc}{ {
	\hypersetup{linkcolor = black} %Locally changes the linkcolor for the toc
	\vspace*{-.06\textheight}	
	\tableofcontents*
	\thispagestyle{empty} 
} }

\makeatletter
\newcommand*\ifthispageodd{%
  \checkoddpage
  \ifoddpage
    \expandafter\@firstoftwo
  \else
    \expandafter\@secondoftwo
  \fi
}
\makeatother

\usepackage[noindentafter]{titlesec} %Removes the indentation after new section
\counterwithout{section}{chapter} % Include if sections should be taken at the top level
\counterwithout{figure}{chapter} %Removes chapter number on figures
\counterwithout{table}{chapter} %Removes chapter number on tables
\numberwithin{equation}{section} % Sets numbering at the section level
\setsecnumdepth{subsubsection}

\DeclareMathVersion{sans}
\SetSymbolFont{operators}{sans}{OT1}{cmbr}{m}{n}
\SetSymbolFont{letters}{sans}{OML}{cmbrm}{m}{it}
\SetSymbolFont{symbols}{sans}{OMS}{cmbrs}{m}{n}
\SetMathAlphabet{\mathit}{sans}{OT1}{cmbr}{m}{sl}
\SetMathAlphabet{\mathbf}{sans}{OT1}{cmbr}{bx}{n}
\SetMathAlphabet{\mathtt}{sans}{OT1}{cmtl}{m}{n}
\SetSymbolFont{largesymbols}{sans}{OMX}{iwona}{m}{n}

\DeclareMathVersion{boldsans}
\SetSymbolFont{operators}{boldsans}{OT1}{cmbr}{b}{n}
\SetSymbolFont{letters}{boldsans}{OML}{cmbrm}{b}{it}
\SetSymbolFont{symbols}{boldsans}{OMS}{cmbrs}{b}{n}
\SetMathAlphabet{\mathit}{boldsans}{OT1}{cmbr}{b}{sl}
\SetMathAlphabet{\mathbf}{boldsans}{OT1}{cmbr}{bx}{n}
\SetMathAlphabet{\mathtt}{boldsans}{OT1}{cmtl}{b}{n}
\SetSymbolFont{largesymbols}{boldsans}{OMX}{iwona}{bx}{n}

\titleformat{\section}{\centering \needspace{5\baselineskip}\Large \bfseries \sffamily \mathversion{boldsans} \color{blue!80!black} }{\thesection}{15pt}{}{}
\titlespacing{\section}{0pt}{15pt}{5pt}
\titleformat{\subsection}{\needspace{2\baselineskip} \large \sffamily \mathversion{sans} \color{blue!70!black} }{\thesubsection}{10pt}{}{}
\titlespacing{\subsection}{0pt}{10pt}{5pt}
\titleformat{\subsubsection}{\normalsize \sffamily \itshape \mathversion{sans} \color{blue!70!black} }{\thesubsubsection}{10pt}{}{}
\titlespacing{\subsubsection}{0pt}{10pt}{0pt}

\newcommand{\sectionlike}[1]{\phantomsection \addcontentsline{toc}{section}{#1} \setcounter{subsection}{0} \sectionmark{#1}
		\begin{center}
		\needspace{5\baselineskip}
		\Large \bfseries \sffamily \mathversion{boldsans} \color{blue!80!black} #1  
		\end{center}
	\vspace{-5pt} 
}

\renewcommand{\paragraph}[1]{\vspace{.3em} \indent {\bfseries \boldmath #1 ---}\xspace }

\ifnum\fullheadfoot=1
	\makepagestyle{standardstyle}
	\makeoddhead{standardstyle}{\sffamily \mathversion{subsectionmath} \rightmark}{}{\sffamily \bfseries{\thepage}}
	\makeevenhead{standardstyle}{\sffamily \bfseries{\thepage}}{}{\sffamily \mathversion{subsectionmath} \leftmark}
	\makeheadrule{standardstyle}{\textwidth}{\normalrulethickness}
	\makepsmarks{standardstyle}{
		\nouppercaseheads
		\createmark{section}{both}{shownumber}{\ }{\hspace{1.2em}  }
		\createmark{subsection}{right}{shownumber}{}{\hspace{1.2em} }
	}
	
	\copypagestyle{appstyle}{standardstyle}
	\makeevenhead{appstyle}{\sffamily \bfseries{\thepage}}{}{ \sffamily \mathversion{subsectionmath} \leftmark}
	\makepsmarks{appstyle}{
		\nouppercaseheads
		\createmark{section}{both}{nonumber}{\ }{\ \ }
		\createmark{subsection}{right}{shownumber}{}{\ \ }
	}
\else %Plain style
	\makepagestyle{standardstyle}
	\makeoddfoot{standardstyle}{}{\sffamily -- {\bfseries\thepage} --}{} 
	\makeevenfoot{standardstyle}{}{\sffamily -- {\bfseries\thepage} --}{}
	\copypagestyle{appstyle}{standardstyle}
\fi

\createplainmark{toc}{both}{\contentsname}
\createplainmark{bib}{both}{\bibname}

\makeatletter
\let\MyIntOrig\int
\def\MyIntSpace{\hspace{-.35em}} %% Configure as needed.
\def\int{\MyInt}
\def\MyInt{\MyIntOrig\MyIntSkipMaybe}
\def\MyIntSkipMaybe{
	\@ifnextchar_{\MyIntSkipScript}{%
		\@ifnextchar^{\MyIntSkipScript}{%
			\@ifnextchar\limits{\MyIntSkipTok}{%
				\@ifnextchar\nolimits{\MyIntSkipTok}{%
					\MyIntSpace}}}}%
}
\def\MyIntSkipScript#1#2{#1{#2}\MyIntSkipMaybe}
\def\MyIntSkipTok#1{#1\MyIntSkipMaybe}

\newcommand{\pushright}[1]{\ifmeasuring@#1\else\omit\hfill$\displaystyle#1$\fi\ignorespaces}
\makeatother

\newcommand{\tr}{\mathop{\mathrm{tr}} }
\newcommand{\eminus}{\vcenter{\hbox{\scalebox{0.6}[1]{$ - $}}}}	%Narrow minus signed (for e.g. negative exponents)

\newcommand{\commutator}[2]{\big[#1, \, #2 \big]}

\newcommand{\dd}{\mathop{}\!\mathrm{d}}

\newcommand{\diag}{\mathop{\mathrm{diag}}}

\newcommand{\sscript}[1]{{\scriptscriptstyle \mathrm{#1}}}

\newcommand{\EFT}{\sscript{EFT}}

\newcolumntype{Y}{>{\centering\arraybackslash}X}
\newcommand{\sumint}[1]{{\hbox{\Large  $\sum$}\!\!\!\!\!\!\!\!\int\,}_{\!\!\raise-0.2ex\hbox{$\scriptstyle{#1}$}}}

\newcommand{\nmode}{{\scriptscriptstyle (n)}}
\newcommand{\zeromode}{{\scriptscriptstyle (0)}}
\newcommand{\nonzeromode}{{\scriptscriptstyle (n\neq 0)}}

\usepackage{accents}
\usepackage{trimclip}

\usepackage{arydshln}
\makeatletter
  \renewcommand*\env@matrix[1][*\c@MaxMatrixCols c]{%
    \hskip -\arraycolsep
    \let\@ifnextchar\new@ifnextchar
  \array{#1}}
\makeatother

\usepackage{graphicx}

\usepackage{mmacells}

\usepackage{etoolbox}

\AtBeginEnvironment{mmaCell}{\fontsize{9.5}{11.5}\selectfont}

\mmaSet{
  morefv={gobble=2},
  linklocaluri=mma/symbol/definition:#1,
  morecellgraphics={yoffset=1.9ex},
  moredefined={Match, DRMatch,SuperSoftMatch, ThermalPowerCounting, LoadModel, NiceForm, CD, 
  DefineField, DefineCoupling, 
  Scalar, Indices, fund, Charges, SelfConjugate,
  SU2L, U1Y, Flavor,
  Mass, Light,
  EFTOrder, LoopOrder, GreensSimplify, EOMSimplify,
  CG, Bar, FreeLag, PlusHc,
  DefineGaugeGroup, SU,Fermion, W0, SelectOperatorClass,LDR,
  SeriesEFT, SeriesThermalEFT, SetDimensions, T, L
  }
}

\newcolumntype{Y}{>{\centering\arraybackslash}X}

\usepackage{tikz}
\usetikzlibrary{positioning,shapes,calc,arrows.meta}
\usepackage{xcolor}

\begin{document}

% % % % Title % % % % 
\thispagestyle{empty}
\renewcommand*{\thefootnote}{\fnsymbol{footnote}}

\begin{center}
%Title
    {\sffamily \bfseries \fontsize{22}{22}\selectfont \mathversion{boldsans}
    Matchotter: An Automated Tool for Dimensional Reduction at Finite Temperature\\[-.005\textheight]
    \textcolor{blue!80!black}{\rule{.5\textwidth}{.7pt}}\\[.015\textheight]}
%Authors
    {\sffamily \mathversion{sans} \Large 
    Javier Fuentes-Martín,\footnote{javier.fuentes@ugr.es}
    Javier López Miras,\footnote{jlmiras@ugr.es} \\ 
    and Adrián Moreno-Sánchez\footnote{adri@ugr.es}
    }\\[1.25em]
%Affilitations
    { \small \sffamily \mathversion{sans} 
        Departamento de Física Teórica y del Cosmos, Universidad de Granada,\\
        Campus de Fuentenueva, E–18071 Granada, Spain
    }
    \\[.005\textheight]{\itshape \sffamily \today}
    \\[.01\textheight]
    \textcolor{blue!80!black}{\rule{.5\textwidth}{.7pt}}\\[.01\textheight]
\end{center}
\setcounter{footnote}{0}
\renewcommand*{\thefootnote}{\arabic{footnote}}%
\suppressfloats	%Prevents figures and the likes on this page 

% % % % Abstract % % % % 

\begin{abstract}\vspace{+.01\textheight}
At finite temperature, the decoupling of heavy Matsubara modes allows a four-dimensional quantum field theory to be matched onto a purely spatial, three-dimensional effective field theory (EFT). This dimensional reduction is a crucial prerequisite for the precise computation of thermal observables, most prominently those related to cosmological phase transitions. In this work, we present \texttt{Matchotter}---a dedicated finite-temperature module natively integrated into the \texttt{Matchete} package~\cite{Fuentes-Martin:2022jrf}---which automates this matching process up to one-loop order for generic Lagrangians. By adapting modern functional matching techniques to the finite-temperature formalism, \texttt{Matchotter} efficiently extracts the low-energy EFT directly from the thermal path integral. Furthermore, the module fully automates supersoft matching, where the temporal gauge bosons, which acquire a Debye mass during the dimensional reduction process, are integrated out. We outline the underlying architecture of the program and demonstrate its capabilities across a range of models, including the Standard Model Effective Field Theory (SMEFT).
\end{abstract}

\newpage
\section*{Table of Contents}
\toc
%\newpage

% % % % % % % % % % % % % % % % % % % % % % % % % % % % % % % %
% % % % % % % % % % % % % % % % % % % % % % % % % % % % % % % %
% % % % % % % % % % % % % % % % % % % % % % % % % % % % % % % %

%%%%%%%%%%%%%%%%%%%%%%%%%%%%%%%%%%%%%%%%%%%%%%%%%%%%%%%%%%%%%%%%%%%%%%%%%%%%%%%%%%%%%%%%%%%%%%
\section{Introduction}
%%%%%%%%%%%%%%%%%%%%%%%%%%%%%%%%%%%%%%%%%%%%%%%%%%%%%%%%%%%%%%%%%%%%%%%%%%%%%%%%%%%%%%%%%%%%%%

Future gravitational-wave observatories and cosmological surveys are actively targeting the observation and characterization of cosmological phase transitions (PTs) as a highly promising window into physics beyond the Standard Model (SM)~\cite{Harry_2006,Kawamura:2006up,Ruan:2018tsw,LISACosmologyWorkingGroup:2022jok}. Because the SM does not predict a first-order PT in either the electroweak or QCD sectors~\cite{Kajantie:1996mn,Csikor:1998eu}, any observation of such a transition would provide definitive evidence of new dynamics. Furthermore, the stochastic gravitational-wave background generated by these transitions carries the unique potential to probe extremely high new physics scales, far exceeding the energy reach of current and foreseeable terrestrial colliders. Consequently, the detection of a gravitational-wave background consistent with a first-order PT constitutes a striking and highly sought-after signature of beyond-the-SM physics (see~\cite{LISACosmologyWorkingGroup:2022jok, Caprini:2019egz} for general reviews).
% ~\cite{Brauner:2016fla,Gould:2019qek,Schicho:2021gca,Biondini:2022ggt,Gould:2022ran,Niemi:2024axp,Annala:2025aci}.

The theoretical framework for studying a quantum field theory immersed in a hot plasma is Thermal Quantum Field Theory (TQFT). Computations within TQFT are typically performed using either the imaginary-time (Matsubara) or the real-time formalism. We adopt the former, which naturally avoids the technical complexities of the real-time formalism and offers a much simpler description for evaluating steady-state properties of PTs. In the Matsubara approach, the presence of a finite temperature $T$ explicitly breaks the (1+3)-dimensional Lorentz symmetry, compactifying the time direction into a periodic domain while leaving the 3-dimensional (Euclidean) space intact. Due to the (anti)periodicity of the boundary conditions for bosonic (fermionic) fields, the fundamental degrees of freedom are decomposed into an infinite tower of Fourier modes, known as Matsubara modes. Crucially, with the sole exception of the \emph{static sector}---composed of bosonic zero modes---all Matsubara excitations acquire a thermal mass proportional to $T$ \cite{Laine:2016hma, Matsubara:1955ws}.

The analysis of PTs within this formalism is naturally suited to an Effective Field Theory (EFT) framework, achieved through a matching procedure known as \emph{dimensional reduction}~\cite{Kajantie:1995dw,Braaten:1995cm}. For a sufficiently high temperature, the heavy Matsubara modes are integrated out, yielding a purely spatial EFT for the light static sector. This paradigm greatly streamlines the computation of PT observables while retaining the advantages of EFTs: it provides a local effective description, improves perturbative convergence near the transition, and enables the resummation of large thermal logarithms. Dimensional reduction techniques for TQFT were first implemented in the study of hot QCD~\cite{Braaten:1995cm,Braaten:1994na,Braaten:1995jr,Kajantie:1997tt,Laine:2019uua,Laine:2018lgj,Ghiglieri:2021bom,Navarrete:2024ruu,Gorda:2025cwu}, and have subsequently emerged as the prevailing framework for investigating cosmological PTs~\cite{Chapman:1994vk,Brauner:2016fla,Andersen:2017ika,Niemi:2018asa,Gorda:2018hvi,Kainulainen:2019kyp,Gould:2019qek,Niemi:2020hto,Gould:2021ccf,Gould:2021dzl,Schicho:2021gca,Lofgren:2021ogg,Niemi:2021qvp,Camargo-Molina:2021zgz,Niemi:2022bjg,Ekstedt:2022bff,Ekstedt:2022ceo,Ekstedt:2022zro,Biondini:2022ggt,Gould:2022ran,Schicho:2022wty,Gould:2023jbz,Kierkla:2023von,Chala:2024xll,Niemi:2024axp,Qin:2024idc,Niemi:2024vzw,Camargo-Molina:2024sde,Gould:2024jjt,Chakrabortty:2024wto,Kierkla:2025qyz,Chala:2025oul,Bernardo:2025vkz,Chala:2025xlk,Chala:2025aiz,Chala:2026ctr,Keus:2025ova,Jahedi:2025yjz,Liu:2025ipj,Li:2025kyo,Annala:2025aci,Bhatnagar:2025jhh,Biekotter:2025npc,Chala:2025cya,Liu:2026ask}. Recently, it has been demonstrated that the inclusion of higher-dimensional operators in the dimensionally-reduced EFT can have a substantial impact on the precise determination of PT parameters~\cite{Chala:2024xll,Bernardo:2025vkz}. This realization has fueled a growing interest in pushing dimensional reduction computations to higher precision~\cite{Bernardo:2026whs,Chala:2025oul}. For instance, a complete implementation of this procedure in the electroweak sector of the Standard Model Effective Field Theory (SMEFT) was only very recently achieved at order $\mathcal{O}(g^4)$~\cite{Chala:2025xlk} and partially at order $\mathcal{O}(g^6)$~\cite{Chala:2025aiz}.

The tedious and highly error-prone nature of these matching calculations, particularly at higher orders or when including higher-dimensional operators, strongly motivates the automation of the dimensional reduction procedure. To date, \texttt{DRalgo}~\cite{Ekstedt:2022bff} has been a highly valuable public code that pioneered the automation of dimensional reduction calculations, greatly simplifying the theoretical pipeline for phenomenological PT analyses. However, its current implementation is mostly restricted to renormalizable interactions in both the underlying theory and the resulting static EFT\footnote{After the first version of this work appeared on the arXiv, a new version of \texttt{DRalgo} was released, which extends the dimensional reduction of renormalizable theories to include operators up to dimension six in the resulting EFT~\cite{Bernardo:2026nyq}.}. This leaves a critical gap for new tools capable of systematically incorporating higher-dimensional operators, thereby expanding the reach and precision of automated dimensional reduction.

In the context of zero-temperature EFTs, functional methods have emerged as a highly robust and efficient alternative to traditional diagrammatic techniques for high-precision matching computations~\cite{Henning:2016lyp,Fuentes-Martin:2016uol,Zhang:2016pja,Cohen:2020fcu,Dittmaier:2021fls}. Rather than computing individual amplitudes, these approaches operate directly at the path-integral level, rendering the matching procedure manifestly gauge-covariant and significantly more streamlined. These functional techniques have already been successfully automated in the public \texttt{Matchete} package~\cite{Fuentes-Martin:2022jrf}, enabling fully automated one-loop matching for conventional EFTs. Ongoing efforts are actively extending this formalism to higher-loop orders~\cite{Fuentes-Martin:2023ljp,Fuentes-Martin:2024agf}, having recently been applied to compute the two-loop renormalization of the SMEFT~\cite{Born:2024mgz,Born:2026xkr}.

One of the main goals of this manuscript is to systematically extend these functional methods to high-temperature dimensional reduction. Furthermore, we introduce \texttt{Matchotter}, a powerful extension of the \texttt{Matchete} package that fully automates dimensional reduction through a user-friendly and highly generalized interface. Alongside this manuscript, we release a new version of \texttt{Matchete} featuring this module (available for download at \url{https://matchete.gitlab.io/}). Within this new version, the user simply defines their preferred high-energy Lagrangian, and the software automatically performs the intricate task of integrating out the heavy Matsubara modes to generate the corresponding dimensionally-reduced EFT.

This manuscript is organized as follows. In Section~\ref{sec:TheoryPart}, we introduce the functional formalism adapted to the dimensional reduction procedure, highlighting the primary methodological differences compared to standard, zero-temperature matching, and provide an illustrative example. In Section~\ref{sec:Matchotter}, we detail the architecture and computer implementation of \texttt{Matchotter}, and provide a practical user guide demonstrating its core functionalities. Section~\ref{sec:Conclusions} contains our conclusions and an outlook for future developments. Finally, Appendices~\ref{app:MatchingFormula} and~\ref{app:SumIntegrals} provide further details on the proof of the functional matching formula and the evaluation of thermal loop integrals, respectively.

%%%%%%%%%%%%%%%%%%%%%%%%%%%%%%%%%%%%%%%%%%%%%%%%%%%%%%%%%%%%%%%%%%%%%%%%%%%%%%%%%%%%%%%%%%%%%%
\section{The functional approach to thermal field theory}
\label{sec:TheoryPart}
%%%%%%%%%%%%%%%%%%%%%%%%%%%%%%%%%%%%%%%%%%%%%%%%%%%%%%%%%%%%%%%%%%%%%%%%%%%%%%%%%%%%%%%%%%%%%%

In this section, we present our functional approach to dimensional reduction in TQFT. The theoretical framework closely follows the methodology developed in~\cite{Fuentes-Martin:2024agf}, which we systematically adapt here to accommodate the finite-temperature case. We begin by formally defining the dimensionally-reduced effective action, emphasizing that the EFT Lagrangian can be extracted directly from its hard region. Next, we detail the application of functional methods to explicitly compute the resulting EFT Lagrangian. Finally, to ground the formal theoretical discussion, we provide an illustrative example where these functional techniques are applied to a toy model.

%%%%%%%%%%%%%%%%%%
\subsection{The dimensionally-reduced effective action}
%%%%%%%%%%%%%%%%%%

We consider an arbitrary quantum field theory described by the Lagrangian density $\mathcal{L}(\eta)$, where $ \eta_a(x) $ denotes a generic collection of scalars, fermions, and gauge bosons along with their conjugate fields (see Section 2.4 of~\cite{Fuentes-Martin:2020udw} for details), with $x$ the spacetime coordinate and $a$ a generic internal index.\footnote{In what follows, we use $x$ and $\vec{x}$ to denote spacetime and space coordinates, respectively. Throughout the manuscript, we employ the mostly minus metric, that is, $g_{\mu\nu}=\diag(+,-,-,-)$.}  For a system at finite temperature $T>0$, the most fundamental quantity of interest is the partition function $\mathcal{Z}_T$, which can be represented by the following path integral 
\begin{align}\label{eq:partitionFunction}
\mathcal{Z}_T[\mathcal{J}]=\int_{\eta_a(0,\vec{x})=\zeta_{aa}\eta_a(\beta,\vec{x})} \;[\mathcal D \eta]\; \exp\!\left\{ -\int_{\tau,\vec{x}}\; \big[\mathcal{L}_E(\eta) + \mathcal{J}_a(\tau,\vec{x})\, \eta_a(\tau,\vec{x}) \big] \right\}\,,
\end{align}
where $\mathcal{J}$ is a source for $\eta$ and we used the shorthand notation $ \int_{\tau,\vec{x}}\; = \int_0^{\beta} \,\dd\tau \int\,\dd^d \vec{x} $ with $\beta\equiv 1/T$. To reflect the presence of fields with mixed spin statistics, it is convenient to arrange them into \emph{superfields} with schematic form $\eta=(\eta_B,\eta_F)$, where $\eta_B$ and $\eta_F$ denote the collections of bosonic and fermionic fields, respectively. The signature $ \zeta_{aa} $ is then introduced to keep track of the Grassmannian signs, with $\zeta_{ab}=-1$ when $a$ and $b$ are fermionic degrees of freedom and $\zeta_{ab}=+1$ otherwise.\footnote{The indices of $\zeta_{ab}$ are \emph{not} counted when determining whether Einstein summation takes place.} The space dimension $ d=3 - 2\epsilon $ is used as a regulator, and the partition function is renormalized perturbatively through the inclusion of counterterms to the Lagrangian. Finally, $\mathcal{L}_E[\eta]$ is the Euclidean Lagrangian density, which is related to the (Minkowskian) Lagrangian by 
\begin{align}
\mathcal{L}_E[\eta]\equiv -\mathcal{L}[\eta]\Big|_{t\to -i\tau,\, A_0\to i A_0}\,,
\end{align}
where $\tau$ is the Euclidean time and $A_0$ denotes the temporal component of a generic gauge field. 

It is convenient to use the (anti)periodic condition for the fields in the partition function to decompose them in terms of their Matsubara modes:
\begin{align}\label{eq:MatsubaraDecomp}
\eta_a(\tau,\vec{x})&= \sqrt{T} \sum_{n=-\infty}^\infty\eta_a^\nmode(\vec{x})\,e^{i\omega_{n,a} \tau}\,,
&
\omega_{n,a}&=2\pi T \Big(n+\frac{1-\zeta_{aa}}{4}\Big)\,,
&
n&\in\mathbb{Z}\,.
\end{align}
Here, positive and negative Matsubara modes are not independent. Because the collection $\eta_a$ is constructed to be self-conjugate (containing all fields alongside their conjugates), its modes are related by complex conjugation. Specifically, the negative-frequency modes of a field map to the positive-frequency modes of its conjugate partner: $(\eta_a^{\scriptscriptstyle (n)})^* = \eta_{\bar{a}}^{\scriptscriptstyle (-n)}$ for bosons, and $(\eta_a^{\scriptscriptstyle (n)})^* = \eta_{\bar{a}}^{\scriptscriptstyle (-n-1)}$ for fermions, where $\bar{a}$ denotes the conjugate field index. However, we choose to retain the full sum over $n \in \mathbb{Z}$ for notational simplicity. As is well known, once this decomposition is inserted in the Lagrangian, the integral over Euclidean time is traded for a sum over Matsubara modes, and all but the bosonic zero modes (i.e. $n=0$) acquire a mass proportional to the temperature.\footnote{The sources are chosen to share the same (anti)periodic conditions of their associated fields, and therefore they are also decomposed analogously. After Euclidean-time integration, the source piece of the partition function becomes 
\begin{align}
\int_0^\beta\dd\tau\, \mathcal{J}_a(\tau,\vec{x})\,\eta_a(\tau,\vec{x})=\sum_{n=-\infty}^\infty \mathcal{J}_a^{\scriptscriptstyle (\eminus n)}(\vec{x})\,\eta_a^\nmode(\vec{x})\,,
\end{align}
with $\mathcal{J}_a^\nmode$ denoting the sources associated with the corresponding Matsubara modes.
}

At energies below $\Lambda_T\equiv 4\pi T$, it becomes more convenient to work directly with the EFT action resulting from integrating out the heavy modes. The process of matching the low-energy behavior of the partition function to that of an EFT made of only bosonic zero modes is commonly termed \emph{dimensional reduction}. The off-shell matching of the original theory and the EFT requires that all connected Green's functions of the bosonic zero modes should be the same at low energies. In turn, this implies that the EFT partition function, $\mathcal{Z}^\textrm{EFT}_T[\mathcal{J}_B^\zeromode]$, can be defined from the low-energy limit of the original partition function by
\begin{align}
\mathcal{Z}^\textrm{EFT}_T[\mathcal{J}_B^\zeromode]=\mathcal{Z}_T[\mathcal{J}_B=\widetilde{\mathcal{J}}_B\equiv \sqrt{T}\mathcal{J}_B^\zeromode,\mathcal{J}_F=0]\,,
\end{align}
where the sources associated with heavy degrees of freedom---that is, bosonic non-zero modes and any fermion mode---are set to zero since they cannot be produced on-shell at low energies. This equality between partition functions should be understood term by term in a power series expansion in $\Lambda_T$, truncated to the desired precision. 
Note that this source choice assumes the original Lagrangian contains no fields with (zero-temperature) masses at or above the thermal scale $\Lambda_T$. Particles much heavier than $\Lambda_T$ must be integrated out prior to dimensional reduction (e.g., using standard zero-temperature functional matching), whereas fields with masses comparable to $\Lambda_T$ must be integrated out alongside the massive Matsubara modes. In the latter scenario, the sources of these massive fields should also be set to zero. We focus our current discussion and implementation on the case where all fields are light compared to $\Lambda_T$. The alternative scenario, while formally straightforward, introduces massive thermal propagators that make the evaluation of the loop corrections more challenging.\footnote{To the best of our knowledge, the resulting finite-temperature loop integrals with massive propagators have only been studied for a few bosonic cases in the literature~\cite{Brauner:2016fla, Navarrete:2025yxy}.}

A more convenient formulation of this condition is obtained by rewriting the partition functions in terms of the finite-temperature (quantum) effective action, the generating functional of all one-particle irreducible (1PI) diagrams. This is obtained via a Legendre transformation of the partition functions as
\begin{align}
\begin{aligned}
\Gamma_T[\hat\eta]&=-\ln \mathcal{Z}_T[\widetilde{\mathcal{J}}_B,\mathcal{J}_F=0] - \widetilde{\mathcal{J}}_{B,I}\, \hat\eta_I\,,
&
\hat{\eta}_I &\equiv -\dfrac{\delta \ln\mathcal{Z}_T[\widetilde{\mathcal{J}}_B,\mathcal{J}_F=0]}{\delta \widetilde{\mathcal{J}}_{B,I}}\,,\\
\Gamma_T^\textrm{EFT}[\hat\eta_B^\zeromode]&=-\ln \mathcal{Z}^\textrm{EFT}_T[\mathcal{J}_B^\zeromode] - \int_{\vec{x}}\;\mathcal{J}_{B,a}^\zeromode\, \hat\eta_a^\zeromode\,,
&
\hat{\eta}^\zeromode_a &\equiv -\dfrac{\delta \ln\mathcal{Z}^\textrm{EFT}_T[\mathcal{J}_B^\zeromode]}{\delta \mathcal{J}_{B,a}^\zeromode}\,,
\end{aligned}
\end{align}
where we use DeWitt notation, in which (Euclidean) time and space are included as part of the indices, namely $I=(\tau,\vec{x},a)$, and thus Einstein summation implies also a spacetime integration such that $\widetilde{\mathcal{J}}_{B,I} \hat\eta_I=\int_{\tau,\vec{x}}\;\, \sqrt{T}\mathcal{J}_{B,a}^\zeromode(\vec{x}) \,\hat{\eta}_a(\tau,\vec{x})= \int_{\vec{x}}\;\mathcal{J}_{B,a}^\zeromode(\vec{x}) \,\hat{\eta}_a^\zeromode(\vec{x})$. The background field $\hat{\eta}$ corresponds to the expectation value of $\eta$ in the presence of the source $\mathcal{J}_B^\zeromode$, which is obtained from the solution to the quantum equations of motion (EOMs) in the presence of zero-mode bosonic background fields. The fact that all Lagrangian terms have an overall Matsubara number zero implies that $\hat{\eta}_B^\nonzeromode=0$ and $\hat\eta_F=0$ at all loop orders. 

All in all, the off-shell matching condition in terms of the finite-temperature effective action simply reads
\begin{align}\label{eq:MatchingCondition-offshell}
\Gamma_T^\textrm{EFT}[\hat\eta_B^\zeromode]=\Gamma_T[\hat\eta_B^\zeromode]\,.
\end{align}
While this matching condition is formally exact, using it directly beyond tree level is highly inefficient, as it requires computing loop corrections in both the full theory and the EFT to extract the EFT matching coefficients. To determine the EFT Lagrangian directly without having to perform loop calculations on the EFT side, this expression requires further refinement. Using expansion by regions~\cite{Beneke:1997zp,Jantzen:2011nz}, it was demonstrated in~\cite{Fuentes-Martin:2016uol,Zhang:2016pja,Fuentes-Martin:2024agf} that the EFT Lagrangian can be identified with the `hard' part of the original effective action. The same arguments can be employed to the thermal case yielding
\begin{align}\label{eq:EFTfromHard}
\int\dd^d\vec{x}\,\mathcal{L}_\EFT[\hat\eta^\zeromode] =  \big(\boldsymbol{R}_\mathrm{hard} \Gamma_T \big) [\hat\eta]\,,
\end{align}
where the operator $ \boldsymbol{R}_\mathrm{hard} $  expands all propagators in the limit where all fields and derivatives are considered small. At one loop, the validity of this matching formula is straightforward: because of Matsubara number conservation, soft contributions in the full theory consist entirely of loops of massless particles, which are in one-to-one correspondence with loops in the EFT. A detailed proof is provided in Appendix~\ref{app:MatchingFormula}. At higher orders, however, subtleties regarding the choice of gauge fixing arise~\cite{Thomsen:2024abg}. Rigorously establishing this formula's validity in gauge theories at multi-loop orders falls beyond the scope of this manuscript and is left for future work. 

As we discuss next, the matching formula in~\eqref{eq:EFTfromHard} provides a simple prescription for dimensional reduction using functional methods.

%%%%%%%%%%%%%%%%%%
\subsection{Evaluating the dimensionally-reduced effective action}
\label{subsec:EffectiveActionEvaluation}
%%%%%%%%%%%%%%%%%%

While recent theoretical progress has systematically extended modern functional methods to multi-loop accuracy \cite{Fuentes-Martin:2023ljp,Born:2024mgz,Born:2026xkr}, we restrict our attention here to the leading quantum corrections, paying special attention to the steps that differ from the zero-temperature case. Having established that the one-loop dimensionally-reduced EFT Lagrangian is entirely encoded within the hard part of the effective action, we now turn to its explicit computation.

The evaluation of the path integral in the effective action relies on standard functional techniques that proceed in a manner completely analogous to zero-temperature quantum field theory. At one-loop order, the effective action takes the particularly simple form (see, e.g.,~\cite{Fuentes-Martin:2024agf} for the explicit derivation)
\begin{align}\label{eq:OneLoopEffectiveAction}
\Gamma_T[\hat\eta^\zeromode]&=S_E^\zeromode[\hat\eta]+\,S_E^{\scriptscriptstyle (1)}[\hat\eta]-\frac{1}{2}\,\mathrm{STr}\log\mathcal{Q}[\hat\eta]\,,
&
\mathcal{Q}_{IJ}[\hat\eta]&= \zeta_{JJ}\dfrac{\delta^2 S_E^\zeromode[\eta]}{\delta \eta_I\, \delta \eta_J}\bigg|_{\eta=\hat{\eta}}\,,
\end{align}
with the supertrace over a generic operator $\mathcal{A}$ defined, in index notation, as $\mathrm{STr}\,\mathcal{A}=\zeta_{II}\mathcal{A}_{II}$. Here, $S_E=\int_{\tau,\vec{x}}\;\mathcal{L}_E$ is the Euclidean action, with $S_E^{\scriptscriptstyle (\ell)}$ ($\ell=0,1$) denoting its tree-level and one-loop parts, and $\mathcal{Q}$ is the so-called \emph{fluctuation operator}. The locality of the action ensures that the fluctuation operator can be decomposed as\footnote{The operators $\mathcal{Q}$ and $U$ do not depend on Euclidean time since the background field $\hat{\eta}$ is a function exclusively of zero modes, that is, $\hat{\eta}_B=\sqrt{T}\,\hat{\eta}_B^\zeromode(\vec{x})$, which results from applying $\hat{\eta}_B^\nonzeromode=0$ in \eqref{eq:MatsubaraDecomp}.}
\begin{align}\label{eq:FlucOpCovDelta}
\mathcal{Q}_{IJ} &= Q_{ac}(\vec{x},\,P_x)\, \delta_{cb}(x,y)\,,
&\mathrm{with}&
&\delta_{ab}(x,\, y) \equiv U_{ab} (\vec{x},\, \vec{y})\; \delta(x-y)\,,
\end{align}
where $Q$ is a differential operator composed of background fields and their derivatives, and open covariant derivatives of the form 
\begin{align}\label{eq:CovDerSplitting}
P_x^\mu\equiv i\widehat{D}_x^\mu=\delta^\mu_j P_x^j- \delta^\mu_0\big[\partial_\tau-i\, \sqrt{T}\widehat{A}_0^\zeromode(\vec{x})\big]\,,    
\end{align}
with $\widehat D_x^\mu=\partial_x^\mu-i \widehat A(x)^\mu$ being the background covariant derivative, $j=1,2,3$ a spatial index and the temperature factor arises from the Matsubara decomposition of the gauge field $\widehat A_0$. Because the (anti)periodic boundary conditions of the partition function explicitly break gauge symmetry along the time direction, gauge invariance is now only manifest for the spatial coordinates. To ensure that the fluctuation operator remains covariant under the remaining spatial gauge transformations, the \emph{covariant delta function}, $\delta_{ab}(\vec{x},\vec{y})$, incorporates a \emph{Wilson line}, $U(\vec{x},\, \vec{y})\equiv \mathscr{P} \exp \! \left[ i \! \int_{\vec{x}}^{\vec{y}} \;\dd z_j \widehat{A}^\zeromode_j(\vec{z}) \right]$ with $\mathscr{P}$ the path-ordering operator. This Wilson line has several useful properties and, in particular, one can show that \cite{Barvinsky:1985an,Kuzenko:2003eb}
\begin{align} \label{eq:DonWL}
\widehat D^j_x\,U(\vec{x},\vec{y}) = i \sum_{k=1}^{\infty} \dfrac{(-1)^k}{(k+1)!} (\vec{x}-\vec{y})_{j_1}\dots(\vec{x}-\vec{y})_{j_k} \big[\widehat D_x^{j_1}\dots \widehat D_x^{j_{k-1}} \widehat A^{j_k j}(\vec{x}) \big]\, U(\vec{x},\vec{y})\,,
\end{align} 
such that in the coincidence limit, where $\vec{x}=\vec{y}$ and $U(\vec{x},\vec{x})=1$, any number of covariant derivatives acting on a Wilson line is written in terms of field-strength tensors, $\widehat A^{ij}(\vec{x})=~[\widehat D_x^i,\widehat D_x^j]$, and their covariant derivatives. As we will see, this property is particularly useful when evaluating the supertraces.

To evaluate the logarithm of the fluctuation operator, it is convenient to decompose it into two pieces
\begin{align}\label{eq:Q2XandDelta}
Q_{ab}(\vec{x},\,P_x)=\Delta_{ab}(P_x)+X_{ab}(\vec{x},P_x)\,,
\end{align}
where $\Delta$ corresponds to the inverse covariant propagator, which is given by\footnote{Although four-dimensional gauge invariance is broken, we retain the four-dimensional momentum operator (i.e., $P_x^\mu$ rather than $P_x^j$) in the inverse covariant propagator to simplify the calculations, bearing in mind that $\partial_\tau \hat\eta^\zeromode(\vec{x})=0$ and $\partial_\tau U_{ab}(\vec{x},\vec{y})=0$. For the same reason, despite the breaking of Lorentz covariance, we maintain four-dimensional gamma matrices in the interaction term for as long as possible.}
\begin{align}\label{eq:DeltaTerms}
\Delta_{ab}&=\delta_{\alpha\beta}\times\left\{
\begin{array}{ccc}
P_x^2 & \qquad\quad & \mathrm{(scalar)}\\
\slashed{P}_{\!\!x} && \mathrm{(fermion)}\\
-g^{\mu\nu}P_x^2 && \mathrm{(vector)}\\
\end{array}
\right.\,,
\end{align}
where $\alpha$ and $\beta$ represent the gauge and flavor components of the indices $a$ and $b$ (which includes also Lorentz and Dirac indices appearing after the bracket in the r.h.s.). Therefore, $\delta_{\alpha\beta}$ should be understood as a product of Kronecker deltas acting exclusively on this internal subspace. If present, mass terms---which, as discussed in Section 2.1, are assumed to lie below the thermal scale $\Lambda_T$---are treated as perturbative interactions and therefore are included in $X$. The vector propagator is gauge dependent; here, and in what follows, we work in the \emph{background-field gauge} with $\xi=1$. In the case of a simple group, the corresponding gauge-fixing Lagrangian is given by
\begin{align}
\mathcal{L}_\mathrm{g.f.}= -\dfrac{1}{2g^2} \big( \widehat{D}^\mu A^A_\mu  \big)^2 - \overline{\omega}^A \widehat{D}^\mu \big(\widehat{D}_\mu \omega^A + f^{ABC} A_\mu^B \omega^C \big)\,,
\end{align}
where $g$ and $f^{ABC}$ are the coupling and structure constants of the gauge group, respectively, and $\omega^A$ ($\overline{\omega}^A$) are the Faddeev-Popov (anti)ghost fields. While this choice is compatible with the matching formula~\eqref{eq:EFTfromHard} at one loop, the gauge-fixing prescription requires closer scrutiny at higher orders.

This splitting of the fluctuation operator lets us decompose the supertrace into a more tractable form
\begin{align}
\begin{aligned}
\mathrm{STr} \log \mathcal{Q} &= \mathrm{STr} [(\log Q)\,\delta] = \mathrm{STr} [(\log \Delta)\,\delta] + \mathrm{STr} [\log\! \big(I - \Delta^{\!\eminus 1} X \big)\,\delta]\\
&= \mathrm{STr} [(\log \Delta)\,\delta] - \sum_{n=1}^{\infty} \dfrac{1}{n} \mathrm{STr} \big[(\Delta^{\!\eminus 1} X)^n\,\delta\big]\,.
\end{aligned}
\end{align}
The first and second terms in this expansion are commonly referred to as \emph{log-type} and \emph{power-type} supertraces. In terms of Feynman diagrams, these encode all one-loop contributions with only bosonic zero modes as external particles, since $\hat\eta_B^\nonzeromode=\hat\eta_F=0$. The log-type supertrace encapsulates the subset of these diagrams where a single internal particle emits an arbitrary number of spatial and temporal zero-mode gauge fields. Conversely, the power-type supertrace accounts for diagrams that feature at least one interaction vertex, thus potentially incorporating loops where multiple distinct particles are present.

\paragraph{Power-type supertrace.} Supertraces can be evaluated using standard functional techniques. The trace is written as a double summation in DeWitt indices constrained by a (covariant) delta function. Therefore, for the power-type supertrace we have
\begin{align}
\begin{aligned}
\mathrm{STr} \big[(\Delta X)^n\,\delta\big] &= \zeta_{aa} \int_{\tau,\vec{x}}\;\int_{\tau^\prime,\vec{y}}\;\delta_{ab}(y,x) \big[\Delta^{\!\eminus 1}(P_x) X(\vec{x},P_x)\big]^n_{bc}\,\delta_{ca}(x,y)\,,
\end{aligned}
\end{align}
with $x=(\tau,\vec{x})$ and $y=(\tau^\prime,\vec{y})$. This expression is clearly divergent due to the singularity of the delta functions. To regularize this singularity, it becomes more convenient to rewrite one of the delta functions in terms of its Fourier decomposition\footnote{The loop momentum $K$ depends on the spin statistics of the field due to the differing bosonic and fermionic Matsubara frequencies, see~\eqref{eq:MatsubaraDecomp}. Hereafter, we use the subscript $K$ to denote a generic sum-integration, reserving $[K]$ and $\{K\}$ for specifically bosonic and fermionic loop momenta, respectively.}
\begin{align}
\delta(x-y)=\sumint{K}\,e^{-i K\cdot (x-y)}\equiv T\sum_{n=-\infty}^\infty\int\frac{\dd^d\vec{K}}{(2\pi)^d}\,e^{-i K\cdot (x-y)}\,,
\end{align}
where $K=(K_0\equiv\omega_n,\vec{K})$ and the series, rather than an integral, on the first coordinate is a consequence of the (anti)periodic boundary conditions. In this way, the singularity in spacetime coordinates gets written into a singularity of \emph{sum-integrals} in loop momentum, which can be regularized in the usual way using dimensional regularization with $d= 3-2\epsilon$. Performing this replacement for the second delta function, letting the differential operator act onto the exponential, and evaluating the remaining delta function, one has
\begin{align}\label{eq:PowSTr}
\begin{aligned}
\mathrm{STr} \big[(\Delta X)^n\,\delta\big] &= \zeta_{aa}\,\frac{1}{T}\!\int_{\vec{x}}\;\sumint{K}\, \big[\Delta^{\!\eminus 1}(P_x+K) X(\vec{x},P_x+K)\big]^n_{ab}\,U_{ba}(\vec{x},\vec{y})\big|_{\vec{x}=\vec{y}}\,,
\end{aligned}
\end{align}
where the factor of $1/T$ arises from the integral in Euclidean time and the coincidence limit for the Wilson line results from the application of the delta function. In this expression, the propagator can be replaced by its series expansion in the limit of large loop momentum $K$. For instance, for the fermion propagator, we have
\begin{align}\label{eq:PropExpansion}
\Delta_{\bar\psi\psi}^{\!\eminus 1}(P_x+K)&=\frac{\slashed{P}_x+\slashed{K}}{(P_x+K)^2}=\frac{\slashed{P}_x+\slashed{K}}{K^2} \sum_{n=0}^\infty \,(\eminus 1)^n \left( \dfrac{2K \cdot P_x + P_x^2}{K^2} \right)^{\!\! n} \,,
\end{align}
and analogously for other field types. 

\paragraph{Log-type supertrace.} Analogously to the power-type supertrace, the log-type supertrace can be expressed as
\begin{align}
\mathrm{STr} [(\log \Delta^{\! \eminus 1})\,\delta] &= \zeta_{aa}\,\frac{1}{T}\!\int_{\vec{x}}\;\sumint{K}\, \log\!\big[\Delta^{\! \eminus 1}(P_x+K)\big]_{ab}\,U_{ba}(\vec{x},\vec{y})\big|_{\vec{x}=\vec{y}}\,,
\end{align}
where, as for the propagator, we can expand the logarithm around $K$. This expansion is straightforward for bosons since $K$ and the operator $P_x$ commute, so we use the properties of the logarithm to rewrite it as
\begin{align}\label{eq:logExpansionBos}
\log\!\big[(P_x+K)^2 \big] = \log[K^2] - \sum_{n=1}^\infty \dfrac{(\eminus 1)^n}{n} \left( \dfrac{2K \cdot P_x + P_x^2}{K^2} \right)^{\!\! n}\,.
\end{align}
The leading term is an infinite constant that will cancel against the partition function normalization and can simply be ignored. The same approach cannot be used directly for fermions because $\commutator{\slashed K}{ \slashed{P}_{\!\!x}} \neq 0$. However, we can simply square the propagator such that
\begin{align}\label{eq:logExpansionFer}
\log\!\big[\slashed K+ \slashed{P}_{\!\!x}\big]=\frac{1}{2}\log\!\big[(\slashed K+ \slashed{P}_{\!\!x})^2\big] = \frac{1}{2}\log[K^2] - \frac{1}{2}\sum_{n=1}^\infty \dfrac{(\eminus 1)^n}{n} \left( \dfrac{2K \cdot P_x + \slashed{P}_{\!\!x} \slashed{P}_{\!\!x}}{K^2} \right)^{\!\! n}\,.
\end{align} 
In this expression, one may be tempted to expand the product $\slashed{P}_{\!\!x} \slashed{P}_{\!\!x}$. However, in practice, it is more convenient to retain this form until it hits the Wilson line.

\bigskip
In general, evaluating the effective action is challenging because expansions such as those in \eqref{eq:PropExpansion} and (\ref{eq:logExpansionBos}, \ref{eq:logExpansionFer}) are not guaranteed to converge. However, since we only need to compute the hard part of the effective action to extract the EFT Lagrangian (see \eqref{eq:EFTfromHard}), we can safely commute the sum-integrals with the series in these expansions. As a result, the matching calculation drastically simplifies, and we are left to evaluate strictly vacuum-type sum-integrals.\footnote{The relevant one-loop sum-integrals can be evaluated analytically. For completeness, we compile their expressions in Appendix~\ref{app:SumIntegrals}.} Following sum-integration, the EFT power counting ensures that only a finite number of terms contribute at any given order, explicitly guaranteeing convergence.

%%%%%%%%%%%%%%%%%%
\subsection{An illustrative example}
%%%%%%%%%%%%%%%%%%

We now illustrate the functional approach described above with a specific example. To this end, we consider the following toy model
\begin{align}\label{eq:Toy-Model}
    \mathcal{L}= -\frac{1}{4g^2}W^{\mu\nu}W_{\mu\nu}+(D_\mu \phi^\dagger)( D^\mu \phi)+\bar\psi i\slashed{D}\psi+ c\,(\phi^\dagger_a \,i\!\overleftrightarrow{D}_{\!\!\mu}\,\phi^b)(\bar \psi_b\gamma^\mu \psi^a)\,,
\end{align}
comprising a fermion $\psi$ and a scalar $\phi$, both chosen to transform in the fundamental representation of an $SU(2)$ gauge group with coupling $g$ and gauge boson $W^\mu$. The coupling $c$ is a real Wilson coefficient with dimensions of $E^{-2}$. We use this simplified model for pedagogical clarity to illustrate the essential functional steps, such as deriving the $X$-terms and evaluating the thermal supertraces. This model will be discussed again in Section~\ref{subsec:code} as an example of its implementation in the code.

For illustration, our aim is to reproduce the one-loop contribution of the higher-dimensional interaction to the quartic term $|\hat\phi^\zeromode|^2 (\widehat W_0^\zeromode{}^I)^2$.\footnote{Computing other contributions like Debye masses, or using more realistic theories like the SMEFT, requires the exact same methodology but involves lengthier algebra that is best handled automatically by \texttt{Matchotter}.} First, we determine the $X$ term involving this higher-dimensional interaction, following from~(\ref{eq:OneLoopEffectiveAction},\,\ref{eq:FlucOpCovDelta},\,\ref{eq:Q2XandDelta}). We have
\begin{align}
[X_{\bar{\psi}\psi}(\vec{x})]^b{}_a =c\,(\hat\phi_a^\dagger \,\!\overleftrightarrow{P}_{\!\!\!\mu}\,\hat\phi^b)\, \gamma^\mu=c\,T\,(\hat\phi_a^\zeromode{}^\dagger\overleftrightarrow{P}_{\!\!\!\mu}\,\hat\phi^\zeromode{}^b)\, \gamma^\mu\,,
\end{align}
where, in the second equality, we used the Matsubara decomposition in~\eqref{eq:MatsubaraDecomp} together with the fact that $\hat\phi^\nonzeromode=0$. As we emphasized in the previous section, even though four-dimensional Lorentz and gauge invariance is broken by the boundary conditions, it is more convenient to retain the four-dimensional covariant derivatives and gamma matrices as long as possible.

The quartic interaction we are interested in only receives contributions from the following supertrace
\begin{align}
\Gamma_T\supset \mathrm{STr}[\Delta_{\bar\psi\psi}^{\eminus1}X_{\bar\psi\psi}\,\delta]\,,
\end{align}
where we accounted for a factor of $2$ since the contributions from $\psi$ and $\psi^c$ enter in the same way in this example. Using~\eqref{eq:PowSTr}, this supertrace is evaluated as
\begin{align}\label{eq:TraceExpansion}
\mathrm{STr}[\Delta_{\bar\psi\psi}^{\eminus1}X_{\bar\psi\psi}\,\delta]\big|_\textrm{hard}
&= -\frac{1}{T}\boldsymbol{R}_\mathrm{hard}\!\int_{\vec{x}}\;\sumint{\{K\}}\, \big[\Delta_{\bar\psi\psi}^{\!\eminus 1}(P_x+K) X_{\bar\psi\psi}(\vec{x})\big]^b{}_a\,U_\psi(\vec{x},\vec{y})^a{}_b\big|_{\vec{x}=\vec{y}}\nonumber\\
&=-\int_{\vec{x}}\;\; c\,(\hat\phi_a^\zeromode{}^\dagger\overleftrightarrow{P}_{\!\!\!\mu}\,\hat\phi^\zeromode{}^b) (P_\rho U_\psi)^a{}_b\big|_{x=y}\tr [\gamma^\mu \gamma^\nu]\;\sumint{\{K\}} \frac{1}{K^2}\left(g_{\nu\rho}-2 \frac{K_\nu K_\rho}{K^2}\right)\nonumber\\
&=-\int_{\vec{x}}\;\;\frac{1}{16\pi^2}\frac{c\Lambda_T^2}{6}\,(\hat\phi_a^\zeromode{}^\dagger\,\!\overleftrightarrow{P}_{\!\!0}\,\hat\phi^\zeromode{}^b) (P_0 U_\psi)^a{}_b\big|_{\vec{x}=\vec{y}}\,,
\end{align}
where, in the second line, we used the expansion in~\eqref{eq:PropExpansion} truncated to terms with up to one derivative since higher-order terms only contribute to higher-dimensional operators. As mentioned in the previous section, keeping only the hard part lets us commute the sum-integral with this expansion. In the last line, we used $\widehat D_j U_\psi\big|_{x=y}=0$, as follows from~\eqref{eq:DonWL}, and evaluated the Dirac trace and the fermion sum-integral using the expressions in Appendix~\ref{app:SumIntegrals}.

At this stage, we express $P_0$ in terms of the background gauge bosons. Using~\eqref{eq:CovDerSplitting} and noting that $\widehat A_0^\zeromode= g\,t^I \widehat W_0^\zeromode{}^I$ when acting on fields in the fundamental representation of $SU(2)$ (with generators $t^I$), we obtain
\begin{align}
\begin{aligned}
(\hat\phi_a^\zeromode{}^\dagger\,\!\overleftrightarrow{P}_{\!\!0}\,\hat\phi^\zeromode{}^b)&=ig\,\sqrt{T}\,[(\hat\phi_c^\zeromode{}^\dagger(t^I)^c{}_a\,\hat\phi^\zeromode{}^b)+(\hat\phi_a^\zeromode{}^\dagger(t^I)^b{}_c\,\hat\phi^\zeromode{}^c)] \,\widehat W_0^\zeromode{}^I\,,\\[0.8 ex]
P_0 U_\psi(\vec{x},\vec{y})^a{}_b\big|_{x=y}&=ig\,\sqrt{T}\, (t^J)^a{}_b \widehat W_0^\zeromode{}^J\,,
\end{aligned}
\end{align}
from where it finally follows that
\begin{align}
\boldsymbol{R}_\mathrm{hard} \Gamma_T&\supset\frac{1}{16\pi^2}\int_{\vec{x}}\;\,\frac{cg^2\Lambda_T^2}{6}\, T|\hat\phi^\zeromode|^2 (\widehat W_0^\zeromode{}^I)^2\,,
\end{align}
where we used the anticommutation relation for the $SU(2)$ generators, namely $\{t^I,t^J\}=\delta_{IJ}$. Using~\eqref{eq:EFTfromHard}, we can directly extract the EFT Lagrangian from this result yielding
\begin{align}\label{eq:result-quartic}
\mathcal{L}_\mathrm{EFT}&\supset-\frac{1}{16\pi^2}\,\frac{c\hat{g}^2\Lambda_T^2}{6}\, |\hat\phi^\zeromode|^2 (\widehat W_0^\zeromode{}^I)^2\,,
\end{align}
with the effective coupling for the spatial gauge symmetry defined as $\hat g\equiv ig\,\sqrt{T}$. The imaginary factor is deliberately introduced so that the gauge boson kinetic operator in the 3D EFT retains a standard Minkowski-like form, $-\frac{1}{\hat{g}^2}F_{ij}F_{ij}$. As we will see, this convention is particularly convenient for the automated algebraic simplifications discussed in the next section.

%%%%%%%%%%%%%%%%%%%%%%%%%%%%%%%%%%%%%%%%%%%%%%%%%%%%%%%%%%%%%%%%%%%%%%%%%%%%%%%%%%%%%%%%%%%%%%
\section{An automated tool for dimensional reduction}
\label{sec:Matchotter}
%%%%%%%%%%%%%%%%%%%%%%%%%%%%%%%%%%%%%%%%%%%%%%%%%%%%%%%%%%%%%%%%%%%%%%%%%%%%%%%%%%%%%%%%%%%%%%

In this section, we introduce \texttt{Matchotter}, a public code that automates the dimensional reduction process by implementing the functional formalism described above. This code is built as an extension of \texttt{Matchete}~\cite{Fuentes-Martin:2022jrf}, a \texttt{Mathematica} package designed to perform functional computations for conventional EFT matching. As such, it is available as an integral part of \texttt{Matchete} since v0.5.

Because dimensional reduction in TQFT is fundamentally a specialized type of EFT matching, the core machinery of \texttt{Matchete} provides an ideal foundation. However, the implementation of dimensional reduction requires several key modifications to the existing code, which we summarize below. Following this, we provide a step-by-step practical guide demonstrating the usage of the package, and conclude by describing our validation of the code against established results in the literature.

%%%%%%%%%%%%%%%%%%%%%%%%%%
\subsection{Computer implementation}
%%%%%%%%%%%%%%%%%%%%%%%%%%

\begin{figure}[h]
    \centering
    \scalebox{0.8}{\input{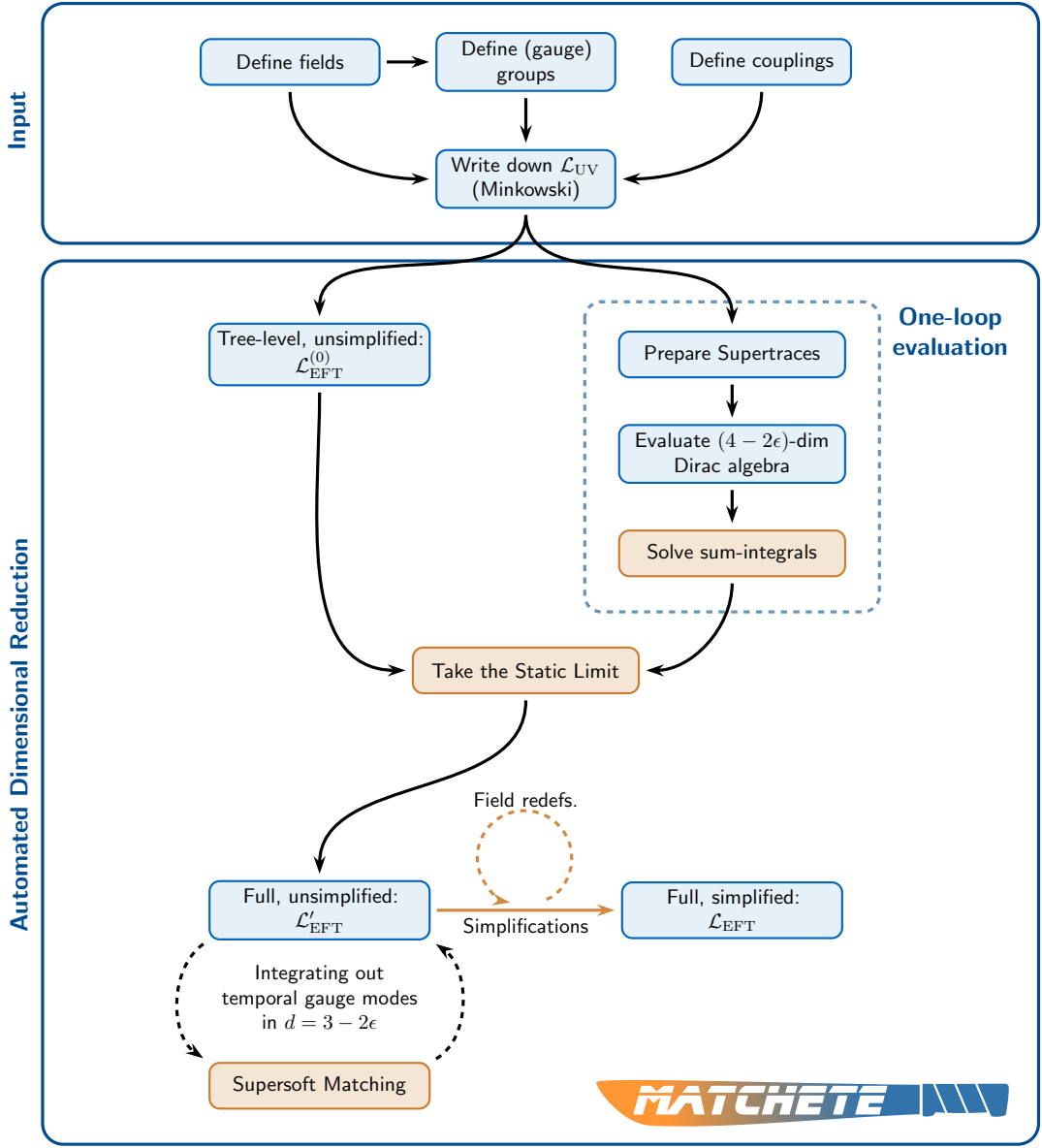}}
    \caption{Workflow of \texttt{Matchotter} for dimensional reduction. Steps highlighted in orange represent specific modifications and new functionalities developed for finite-temperature matching.}
    \label{fig:workflow}
\end{figure}

An illustrative diagram of the \texttt{Matchotter} workflow is shown in Fig.~\ref{fig:workflow}. The definition of the four-dimensional Lagrangian and the construction of the supertraces rely entirely on the standard setup and tools provided by the base \texttt{Matchete} code. Initially, these supertraces preserve explicit Lorentz covariance, facilitating the evaluation of the Dirac traces. However, this covariance is manifestly broken when computing the thermal sum-integrals, a step that requires a Lorentz-breaking tensor reduction in which the temporal dimension is treated separately from the spatial ones. This represents the key technical departure from standard matching computations, leading to sum-integrals that depend on zero and spatial components of the loop momenta, together with four-dimensional propagators. The general expressions and closed formulas for these one-loop vacuum type sum-integrals, hardcoded into \texttt{Matchotter}, are provided in Appendix~\ref{app:SumIntegrals}.

Following the sum-integral evaluation, the resulting expressions factorize into distinct spatial and temporal structures. We further apply this separation to any remaining Lorentz index contractions, which allows us to recursively split all covariant derivatives according to~\eqref{eq:CovDerSplitting}. For computational efficiency, the temporal gauge field components, $\hat{A}^\zeromode_0$, are kept unexpanded at this stage. Next, all spatial derivatives are allowed to act on objects to their right until they hit a Wilson line; following~\eqref{eq:DonWL}, any derivatives acting on a Wilson line are then replaced by field-strength tensors and their derivatives. 

Afterwards, we take what we refer to as the \emph{static limit}. In this step, all fields are replaced by their classical configurations, and the temporal gauge bosons are explicitly expanded. \texttt{Matchotter} automatically defines the zero modes as new fields. By default, their names are generated by appending a ``0'' to the original field labels, though custom names can easily be specified by the user. Furthermore, the package defines new gauge groups associated with the spatial restriction of the gauge symmetry. The names of these new groups, along with their gauge fields and couplings, are formed by appending ``Sp'' to their parent names. At this stage, we also perform various sign adjustments to adapt the Minkowskian \texttt{Matchete} expressions to Euclidean. Finally, we replace any appearances of the original gauge couplings, $g_i$, with their spatial counterparts, $\hat{g}_i$, following the relation $\hat{g}_i = ig_i\,\sqrt{T}$. 

Once the dimensionally-reduced Lagrangian is obtained, the rest of the code workflow relies on the built-in \texttt{Matchete} algorithms to reduce the result to either an off-shell or on-shell basis. However, because the resulting effective theory is purely static, applying these standard routines required targeted adaptations to accommodate static Lagrangians. Specifically, the internal machinery was generalized to handle contractions over spatial rather than Lorentz indices, alongside the proper algebraic evaluation of spatial derivatives and purely spatial Levi-Civita tensors. Beyond these adaptations, the code now also incorporates a thermal power counting. This power counting can be included as part of the coupling definitions, and a dedicated function is provided to systematically expand and truncate the Lagrangian based on this counting (see section~\ref{subsec:code}).

In addition to the primary dimensional reduction workflow, \texttt{Matchotter} also includes a dedicated function for subsequent thermal EFT manipulations. Most notably, we implement the \emph{supersoft} matching, a standard matching procedure within the resulting dimensionally-reduced EFT in which the massive temporal gauge fields are integrated out. Adapting the code for this step primarily required generalizing \texttt{Matchete} to operate in arbitrary dimensions, a byproduct feature that is now available also in the standard \texttt{Matchete} pipeline.

Ultimately, \texttt{Matchotter} allows the user to automatically perform one-loop dimensional reduction and subsequent supersoft matching for a broad class of theories with particles whose masses lie below the thermal scale, systematically expanding the resulting effective Lagrangian up to any desired operator dimension.

%%%%%%%%%%%%%%%%%%%%%%%%%%
\subsection{Validation of the code}
%%%%%%%%%%%%%%%%%%%%%%%%%%

To ensure the robustness and accuracy of the toolkit, it has undergone rigorous validation against several established models in the literature. These models were selected to test different functionalities of the code. Specifically, we have successfully crosschecked the following results:

\begin{itemize}
    \item \textbf{Pure $\boldsymbol{SU(N)}$ Yang-Mills theory}~\cite{Chapman:1994vk}. Although the original authors employed a different regularization scheme (i.e., zeta function regularization), we can meaningfully compare the finite contributions to the effective operators. We observe a disagreement in three redundant operators, which vanishes entirely upon going on-shell and can thus be absorbed by field redefinitions. Such an off-shell discrepancy is likely due to the use of different gauge-fixing schemes.

    \item \textbf{Abelian Higgs model}~\cite{Bernardo:2025vkz}. In addition to the primary dimensional reduction workflow, this reference allowed us to validate our supersoft matching routine. We found complete agreement with the supersoft limit of the EFT derived therein.

    \item \textbf{Yukawa toy model}~\cite{Chala:2024xll}. Although conceptually simple, this model is crucial for verifying the correct handling of fermions and Dirac traces. We successfully crosschecked the resulting effective operators up to mass dimension 8.

    \item \textbf{Electroweak sector of the SMEFT}~\cite{Chala:2025aiz, Chala:2025xlk}. This constitutes our most exhaustive crosscheck. Within this model, \texttt{Matchotter}'s treatment of interactions among abelian and non-abelian gauge fields, scalars, and fermions is thoroughly tested. While we initially identified some discrepancies with~\cite{Chala:2025aiz}, private communication with the authors confirmed that these stemmed from errors and sign inconsistencies in the ancillary files of the original publication. The authors are currently updating these files, and our output shows full agreement with their corrected versions.

    \item \textbf{One-loop Renormalization Group equations}~\cite{Chala:2025crd}. Following the methodology proposed in this reference, we performed a dimensional reduction from 5 to 4 dimensions. This allowed us to successfully reproduce the corresponding subset of the one-loop renormalization group equations. This provides an important consistency check that the dimensional reduction algorithms operate correctly in $d \neq 3$ spatial dimensions.
\end{itemize}

\begin{table}[t]
    \centering
    \begin{tabularx}{\linewidth}{YYYY}
        \toprule
        $SU(3)$ Yang-Mills & Abelian Higgs Model & Yukawa Toy Model & Dim-6 SMEFT
        \\ \midrule
        3.\,56 s & 4.\,66 s & $\sim$3.\,5 min  & $\sim$30 min
        \\
        \bottomrule
    \end{tabularx}
    \caption{Time benchmarks for dimensional reduction using \texttt{Matchotter} across different test models.}
    \label{tab:time benchmarks}
\end{table}

To highlight \texttt{Matchotter}'s performance in terms of execution time, we provide a series of benchmarks in Table~\ref{tab:time benchmarks}. With the exception of the Yukawa toy model (which was expanded up to operator dimension 8), all models were evaluated up to dimension 6. Notably, the dimension-6 SMEFT benchmark uses the full theory as defined in~\cite{Grzadkowski:2010es}, rather than being restricted to the electroweak sector. All tests were carried out on a single machine equipped with an AMD Ryzen 9 CPU and 32 GB of RAM.

%%%%%%%%%%%%%%%%%%%%%%%%%%
\subsection{\texttt{Matchotter} in action}
\label{subsec:code}
%%%%%%%%%%%%%%%%%%%%%%%%%%

In this section, we provide a practical demonstration of \texttt{Matchotter} by explicitly performing the dimensional reduction of the toy model introduced in~\eqref{eq:Toy-Model}. The first step consists of defining the four-dimensional model using the standard \texttt{Matchete} syntax\footnote{For a comprehensive guide on model definitions, we refer the reader to \texttt{Matchete}'s paper~\cite{Fuentes-Martin:2022jrf} or webpage \url{https://matchete.gitlab.io/}.}
\begin{mmaCell}[]{Input}
  DefineGaugeGroup[SU2,\(\,\,\)SU[2],\(\,\,\)g,\(\,\,\)W]
  DefineField[\mmaUnd{\(\phi\)},\(\,\,\)Scalar,\(\,\,\)Indices\(\to\)SU2[fund],\(\,\,\)Mass\(\to\)0]
  DefineField[\mmaUnd{\(\psi\)},\(\,\,\)Fermion,\(\,\,\)Indices\(\to\)SU2[fund],\(\,\,\)Mass\(\to\)0]
  DefineCoupling[c,\(\,\,\)SelfConjugate\(\to\)True]
  
  \mmaUnd{L} = FreeLag[] +
     c[] PlusHc[\(\,\,\)I(Bar[\(\phi\)[a]]CD[\(\mu\),\(\,\,\)\(\phi\)[b]])(Bar[\(\psi\)[b]]\(\,\,\)\(\cdot\)\(\,\,\)\(\gamma\)[\(\mu\)]\(\,\,\)\(\cdot\)\(\,\,\)\(\psi\)[a])]//NiceForm
  
\end{mmaCell}
\begin{mmaCell}[]{OutputNiceForm}
  -\mmaFrac{1}{4\mmaSup{g}{2}}\mmaSup{\mmaSup{W}{\(\mu\nu\)I}}{2}\(\,\,\,\)+\(\,\,\,\)\mmaSub{D}{\(\mu\)}\mmaSub{\(\overline{\phi}\)}{a}\mmaSup{D}{\(\mu\)}\mmaSup{\(\phi\)}{a}\(\,\,\,\)+\(\,\,\,\)i(\mmaSub{\(\overline{\psi}\)}{a}\(\cdot\)\mmaSub{\(\gamma\)}{\(\mu\)}\(\cdot\)\mmaSup{D}{\(\mu\)}\mmaSup{\(\psi\)}{a})\(\,\,\,\)+\(\,\,\,\)i\(\,\)c\(\,\)\mmaSub{\(\overline{\phi}\)}{a}\mmaSup{D}{\(\mu\)}\mmaSup{\(\phi\)}{b}(\mmaSub{\(\overline{\psi}\)}{b}\(\cdot\)\mmaSub{\(\gamma\)}{\(\mu\)}\(\cdot\)\mmaSup{\(\psi\)}{a})\(\,\,\,\)-\(\,\,\,\)i\(\,\)c\(\,\)\mmaSup{D}{\(\mu\)}\mmaSub{\(\overline{\phi}\)}{a}\mmaSup{\(\phi\)}{b}(\mmaSub{\(\overline{\psi}\)}{b}\(\cdot\)\mmaSub{\(\gamma\)}{\(\mu\)}\(\cdot\)\mmaSup{\(\psi\)}{a})

\end{mmaCell}
Once the high-energy Lagrangian is defined, dimensional reduction is executed via a single command\footnote{In \texttt{Matchete} v0.5.0, the \texttt{DRMatch} function was originally named \texttt{DimensionalReduce}. It was renamed in v0.5.1 to avoid confusion with Mathematica's built-in \texttt{DimensionReduce} function.}
\begin{mmaCell}[]{Input}
  \mmaUnd{LDR} = DRMatch[L, EFTOrder\(\to\)6, LoopOrder\(\to\)1];

\end{mmaCell}
The option \mmaInlineCell{Input}{EFTOrder} specifies the maximum mass dimension of the operators retained in the resulting static EFT (set to 6 by default). Similarly, the option \mmaInlineCell{Input}{LoopOrder} indicates the number of loops in the matching calculation. This option is set to 1 by default, which is the maximum value supported in the current version of \texttt{Matchotter}. Although shown here for illustration, both options may be omitted when their default values are desired. The resulting effective Lagrangian can then be reduced to an off-shell or on-shell basis using the standard \texttt{Matchete} functions \mmaInlineCell{Input}{GreensSimplify} and \mmaInlineCell{Input}{EOMSimplify}, respectively. Crucially, the internal mechanics of these simplification routines have been adapted to properly handle dimensionally-reduced, static Lagrangians. 

The output Lagrangian can be manipulated using the standard suite of \texttt{Matchete} tools. For instance, to isolate and verify the quartic interaction derived analytically in~\eqref{eq:result-quartic}, the user can execute\footnote{As explained in Section~\ref{sec:Matchotter}, the resulting expressions are given in terms of zero modes. These are automatically defined by the \texttt{DRMatch} function and assigned the labels \texttt{$\phi$0} and \texttt{W0}. Likewise, the original gauge symmetry and its associated field and coupling are replaced by their spatial counterparts. In this example, the spatial gauge group is named \texttt{SU2Sp}, while its corresponding gauge coupling and gauge field are labeled \texttt{gSp} and \texttt{WSp}, respectively.}
\begin{mmaCell}[]{Input}
  SelectOperatorClass[EOMSimplify[LDR],\(\,\{\phi0,\,\,\phi0,\,\,W0,\,\,W0\}\),\(\,\,\)0]// NiceForm

\end{mmaCell}
\begin{mmaCell}[]{OutputNiceForm}
  \(\Big(\)-\mmaFrac{1}{4}\mmaSup{\(\hat{g}\)}{2}+\(\hbar\)\(\Big(\)-\mmaFrac{1}{6}\mmaSubSup{\(\Lambda\)}{\(\mathcal{T}\)}{2}c\mmaSup{\(\hat{g}\)}{2}+\mmaFrac{55}{24}\mmaFrac{1}{\(\mathcal{T}\)}\mmaSup{\(\hat{g}\)}{4}+\mmaFrac{13}{18}\mmaFrac{1}{\(\epsilon\)}\mmaFrac{1}{\(\mathcal{T}\)}\mmaSup{\(\hat{g}\)}{4}+\mmaFrac{43}{24}\mmaFrac{1}{\(\mathcal{T}\)}\mmaSup{\(\hat{g}\)}{4}Log\(\Big[\)\mmaFrac{\mmaSup{\(\overline{\Lambda}\)}{2}}{\mmaSubSup{\(\Lambda\)}{\(\mathcal{T}\)}{2}}\(\Big]\)-\mmaFrac{1}{6}\mmaFrac{1}{\(\mathcal{T}\)}\mmaSup{\(\hat{g}\)}{4}Log\(\Big[\)\mmaFrac{16\mmaSup{\(\overline{\Lambda}\)}{2}}{\mmaSubSup{\(\Lambda\)}{\(\mathcal{T}\)}{2}}\(\Big]\)\(\Big)\)\(\Big)\)\mmaSub{\(\overline{\phi}\)}{0a}\mmaSubSup{\(\phi\)}{0}{a}\mmaSubSup{W}{0}{I\(\,\)2}

\end{mmaCell}
Here, we applied an on-shell reduction of the Lagrangian prior to selecting the operator class---operators with two powers of $\widehat \phi^\zeromode$ and two powers of $\widehat W^\zeromode$---with the \mmaInlineCell{Input}{SelectOperatorClass} routine. The second term in this output exactly matches the analytical expression in~\eqref{eq:result-quartic}, noting that \texttt{Matchete} employs the convention \mmaInlineCell{Input}{\(\hbar\)} $= 1/(16\pi^2)$. The additional terms present in this output arise once all contributions---and not only those proportional to $c$---are considered.

Beyond the static limit, the user can seamlessly proceed to the \emph{supersoft matching} phase. This step involves integrating out the temporal gauge bosons---which acquire Debye masses after dimensional reduction---to match onto an EFT valid below the supersoft scale. Rather than requiring the user to manually redefine the heavy states and perform a standard matching procedure, \texttt{Matchotter} provides the dedicated \mmaInlineCell{Input}{SuperSoftMatch} function. This routine automatically takes the dimensionally-reduced Lagrangian as input, identifies the temporal gauge components and their thermal masses, and integrates them out.
\begin{mmaCell}[]{Input}
  SuperSoftMatch[LDR, EFTOrder\(\to\)6, LoopOrder\(\to\)1, ThermalPowerCounting\(\to\)6]

\end{mmaCell}
In this command, we have introduced the \mmaInlineCell{Input}{ThermalPowerCounting} option, which systematically truncates the resulting EFT according to an expansion in powers of a generic coupling $\lambda$. By default, \texttt{Matchotter} assumes that every coupling in the UV theory scales linearly with this coupling ($\lambda_i \sim \lambda^1$). However, this scaling behavior can be customized by the user during the initial model definition. For example, if a specific coupling $c$ scales as $c \sim \lambda^2$, it should be defined as
\begin{mmaCell}[]{Input}
  DefineCoupling[c,\(\,\,\)SelfConjugate\(\to\)True, ThermalPowerCounting\(\to\)2]
\end{mmaCell}
This thermal truncation can also be applied manually to any standalone expression using \mmaInlineCell{Input}{SeriesThermalEFT}, a thermally-adapted counterpart to the standard \mmaInlineCell{Input}{SeriesEFT} function.

Finally, while these tools default to four spacetime dimensions, \texttt{Matchotter} allows for dimensional reduction in arbitrary dimensions, a feature that may prove useful in certain applications~\cite{Chala:2025crd}. The spacetime dimensionality can be dynamically adjusted using the \mmaInlineCell{Input}{SetDimensions} command, e.g.,
\begin{mmaCell}[]{Input}
   SetDimensions[5]
\end{mmaCell}
This function changes the number of dimensions globally, affecting any other routine for which the number of dimensions is a relevant input, such as the standard \mmaInlineCell{Input}{Match} function.

%%%%%%%%%%%%%%%%%%%%%%%%%%%%%%%%%%%%%%%%%%%%%%%%%%%%%%%%%%%%%%%%%%%%%%%%%%%%%%%%%%%%%%%%%%%%%%
\section{Conclusions and outlook}
\label{sec:Conclusions}
%%%%%%%%%%%%%%%%%%%%%%%%%%%%%%%%%%%%%%%%%%%%%%%%%%%%%%%%%%%%%%%%%%%%%%%%%%%%%%%%%%%%%%%%%%%%%%

The study of cosmological phase transitions relies on the precise matching of high-temperature theories onto their dimensionally-reduced EFT counterparts via a process known as dimensional reduction. However, performing dimensional reduction without a fully automated pipeline is a laborious and highly error-prone endeavor, representing a major bottleneck for phenomenological explorations of early-universe physics. To address this, the present work makes a twofold contribution: developing the theoretical formalism required to systematize finite-temperature matching via functional methods, and providing a public software tool that fully automates it.

In the first part of this paper, we established the necessary theoretical groundwork by adapting modern functional methods to the finite-temperature regime. This formalism bypasses the inefficiencies of traditional diagrammatic approaches, such as the lack of manifest gauge covariance at intermediate stages of the computation or the need to predetermine the target EFT. Building directly upon this formalism, we introduced \texttt{Matchotter}, a dedicated finite-temperature module natively integrated into the \texttt{Matchete} package. \texttt{Matchotter} automates the dimensional reduction of generic four-dimensional theories (including EFTs) onto purely spatial EFTs at the one-loop level. Furthermore, \texttt{Matchotter} automates supersoft matching, systematically integrating out the temporal gauge bosons---which acquire an intermediate Debye mass---to construct an EFT valid at energy scales below this threshold. The module has been thoroughly validated against a diverse set of benchmark models, ranging from simplified toy models to the electroweak sector of the SMEFT. 

Looking ahead, achieving theoretical consistency in thermal power counting demands pushing these matching computations to higher loop orders. Different sectors of the effective Lagrangian inherently require matching at higher loops simply to maintain a uniform level of precision across the full theory. Consequently, the most critical future extension for \texttt{Matchotter} is the generalization of the dimensional reduction process to two loops, and potentially to higher orders for specific contributions. The primary bottleneck preventing this extension is theoretical: extending the functional formalism to this level requires a careful treatment of the gauge-fixing terms, as an explicit proof of the hard-region matching formula~\eqref{eq:EFTfromHard} is central for any rigorous functional evaluation. Once these theoretical aspects are clarified, the finite-temperature machinery must be systematically adapted to interface with the two-loop zero-temperature extensions currently being developed for the core \texttt{Matchete} framework~\cite{Fuentes-Martin:2024agf,Born:2024mgz,Born:2026xkr}.\footnote{Although conceptually solved, the systematic evaluation of two-loop massless vacuum sum-integrals may pose another potential technical difficulty. However, they have been proved to factorize into products of one-loop sum-integrals via integration by parts (IBP) relations~\cite{Nishimura:2012ee, Ghisoiu:2012yk, Moller:2012chx} and a complete factorization formula for the purely bosonic case is demonstrated in~\cite{Davydychev:2023jto}. Crucially, a public code performing automatic IBP reduction for sum-integrals will be soon published~\cite{Gil:2026cqz}.} While this integration will present its own computational challenges, the structure of the package is flexible enough to accommodate it, just as we have demonstrated at one-loop order.

Ultimately, we anticipate that the public release of \texttt{Matchotter} as an integral part of the \texttt{Matchete} ecosystem, alongside the theoretical formalism it implements, will significantly alleviate the computational burden associated with finite-temperature EFTs. By providing a robust, unified, and automated pipeline, this tool empowers the community to rapidly and reliably explore the cosmological implications of a wide landscape of beyond-the-Standard-Model theories.

%%%%%%%%%%%%%%%%%%%%%%%%%%%%%%%%%%%%%%%%%%%%%%%%%%%%%%%%%%%%%%%%%%%%%%%%%%%%%%%%%%%%%%%%%%%%%%
\subsection*{Acknowledgments}
%%%%%%%%%%%%%%%%%%%%%%%%%%%%%%%%%%%%%%%%%%%%%%%%%%%%%%%%%%%%%%%%%%%%%%%%%%%%%%%%%%%%%%%%%%%%%%

We are grateful to Mikael Chala, Guilherme Guedes, Anders E. Thomsen, and Felix Wilsch for useful comments on the manuscript and to Mikael Chala and Guilherme Guedes for their help in comparing our results with those in~\cite{Chala:2025aiz}. 
% Funding: all
This work is supported by the European Research Council under grant agreement n. 101230200 and the Spanish Research Agency (MICIU/AEI/10.13039/501100011033) and the European Union (FEDER/UE) under grant PID2022-139466NB-C21, as well as by the Junta de Andaluc\'ia grant FQM101. 
% Funding: JFM and AMS
The work of JFM and AMS is further supported by grants CNS2024-154834, EUR2024.153549 by the Spanish Research Agency (MICIU/AEI/10.13039/501100011033) and the European Union (NextGenerationEU/PRTR and FEDER/UE). 
% Funding: AMS and JLM
Furthermore, the work of AMS and JLM is supported by the Spanish government and the European Union – NextGenerationEU through grants FPU23/01639 and FPU23/02028, respectively.

\renewcommand{\thesection}{\Alph{section}}
\appendix 

%%%%%%%%%%%%%%%%%%%%%%%%%%%%%%%%%%%%%%%%%%%%%%%%%%%%%%%%%%%%%%%%%%%%%%%%%%%%%%%%%%%%%%%%%%%%%%
%%%%%%%%%%%%%%%%%%%%%%%%%%%%%%%%%%%%%%%%%%%%%%%%%%%%%%%%%%%%%%%%%%%%%%%%%%%%%%%%%%%%%%%%%%%%%%

%%%%%%%%%%%%%%%%%%%%%%%%%%%%%%%%%%%%%%%%%%%%%%%%%%%%%%%%%%%%%%%%%%%%%%%%%%%%%%%%%%%%%%%%%%%%%%
\section{Master Matching Formula}
\label{app:MatchingFormula}
%%%%%%%%%%%%%%%%%%%%%%%%%%%%%%%%%%%%%%%%%%%%%%%%%%%%%%%%%%%%%%%%%%%%%%%%%%%%%%%%%%%%%%%%%%%%%%

In this section, we prove that the EFT Lagrangian can be obtained directly from the hard part of the effective action of the UV theory, $\Gamma_T$. We begin with the off-shell matching condition in~\eqref{eq:MatchingCondition-offshell}. From this equation, we can directly extract the EFT action at tree level,
\begin{align}\label{eq:treelevel_matching}
    \mathcal{S}^\zeromode_{EFT}[\hat\eta_B^\zeromode]= \mathcal{S}^\zeromode_{E}[\hat\eta]\,,
\end{align}
where, as a reminder, $\hat\eta=(\hat\eta_B=\sqrt{T}\,\hat\eta_B^\zeromode,\,\hat\eta_F=0)$, with $\eta_B$ and $\eta_F$ used to generically denote bosonic and fermionic degrees of freedom, respectively. Beyond tree level, this relation becomes more subtle. First, we expand the effective action up to one-loop order as
\begin{align}\label{eq:QET_oneloop}
\Gamma_T[\hat\eta]&=S_E^\zeromode[\hat\eta]+\,S_E^{\scriptscriptstyle (1)}[\hat\eta]-\frac{1}{2}\,\mathrm{STr}\log\mathcal{Q}[\hat\eta]+\,\mathcal{O}(\hbar^2)\,,
\end{align}
with $\mathcal{Q}$ being the so-called fluctuation operator, which is given by
\begin{align}\label{eq:fluctuationoperator}
    \mathcal{Q}_{I^\nmode J^{\scriptscriptstyle (m)}}[\hat\eta]=\zeta_{JJ}\frac{\delta \mathcal{S}^\zeromode_E[\eta]}{\delta\eta^\nmode_I\delta\eta^{\scriptscriptstyle (m)}_J}\Bigg|_{\eta=\hat\eta}\,.
\end{align}
Here, we have made the Matsubara number of the modes explicit, as it will prove useful in what follows. Indeed, when this operator is evaluated at the background configurations, it becomes diagonal in the Matsubara decomposition. This property allows us to split the supertrace in~\eqref{eq:QET_oneloop} into two contributions:
\begin{align}\label{eq:fluctuationoperatorDecomposition}
    \mathrm{STr}\log\mathcal{Q}[\hat\eta]=(\log\mathcal{Q}[\hat\eta_B^\zeromode])_{I^\zeromode I^\zeromode}+\sum_{n\neq0}(\log\mathcal{Q}[\hat\eta_B^\zeromode])_{I^\nmode I^\nmode}\,.
\end{align}
This splitting naturally separates the supertrace into a soft contribution (involving exclusively loops with zero modes) and a hard contribution (comprising loops with exclusively non-zero modes, whose masses lie at or above the heavy scale $\Lambda_T$). This straightforward separation is due to the absence of heavy-light loops, a special feature of thermal dimensional reduction arising from Matsubara number conservation. Furthermore, using the tree-level EFT action~\eqref{eq:treelevel_matching}, one can readily see that the corresponding fluctuation operator in the EFT gives the same contribution as the first term in~\eqref{eq:fluctuationoperatorDecomposition}, such that
\begin{align}
\Gamma_T^\textrm{EFT}[\hat\eta_B^\zeromode]&=S_E^\zeromode[\hat\eta]-\frac{1}{2}\,(\mathrm{STr}\log\mathcal{Q}[\hat\eta_B^\zeromode])_{I^\zeromode I^\zeromode}+\,\mathcal{O}(\hbar^2)\,.
\end{align}
This establishes a one-to-one correspondence between loops involving only zero modes in the UV theory and loops in the EFT. The remaining contribution therefore corresponds to the one-loop matching, yielding
\begin{align}
    \mathcal{S}^{\scriptscriptstyle (1)}_\textrm{EFT}[\hat\eta_B^{(0)}]=\Gamma_T[\hat\eta]-\Gamma_T^\textrm{EFT}[\hat\eta_B^\zeromode]=S_E^{\scriptscriptstyle (1)}[\hat\eta]-\frac{1}{2}\,\sum_{n\neq0}(\log\mathcal{Q}[\hat\eta_B^\zeromode]) _{I^{(n)}I^{(n)}}\,.
\end{align}
This result coincides exactly with the hard part of the effective action of the UV theory, resulting in the resulting expression given in~\eqref{eq:EFTfromHard}.

Beyond one loop, this cancellation of loop contributions is no longer straightforward, since the loops generally contain a mixture of heavy and light propagators. For theories free of gauge bosons, the general arguments in~\cite{Fuentes-Martin:2024agf} can be readily adapted to obtain a valid thermal matching formula at multi-loop order. The more interesting case in which gauge bosons are present, however, remains more subtle, especially for certain gauge-fixing conditions~\cite{Thomsen:2024abg}. The extension of the matching formula to multi-loop order lies beyond the scope of the present work and will be addressed in future studies.

%%%%%%%%%%%%%%%%%%%%%%%%%%%%%%%%%%%%%%%%%%%%%%%%%%%%%%%%%%%%%%%%%%%%%%%%%%%%%%%%%%%%%%%%%%%%%%
\section{Analytic expressions for the sum-integrals}
\label{app:SumIntegrals}
%%%%%%%%%%%%%%%%%%%%%%%%%%%%%%%%%%%%%%%%%%%%%%%%%%%%%%%%%%%%%%%%%%%%%%%%%%%%%%%%%%%%%%%%%%%%%%

This appendix introduces the main ingredients for the analytic evaluation of the sum-integrals appearing during dimensional reduction process at one-loop order. As mentioned in Section~\ref{sec:TheoryPart}, we only need to consider vacuum-type (tensorial) sum-integrals, which take the general form
\begin{align}
I^m_{\mu_1,\dots,\mu_{2n}}= \sumint{K} \frac{K_{\mu_1}\dots K_{\mu_{2n}}}{K^{2m}}\, ,
\end{align}
where $K$ is a generic loop momentum, which can be bosonic or fermionic. The number of Lorentz indices must be even for the sum-integral to be non-vanishing in dimensional regularization, and these indices are contracted with metric factors or with other operator structures from the supertrace.

It is convenient to apply tensor reduction identities to rewrite these sum-integrals in terms of scalar ones. As sum-integrals explicitly break the Lorentz symmetry, these take the following form  \cite{Navarrete:2022adz}
\begin{align}\label{eq:TesorRedSumInt}
\sumint{K} \frac{K_{\mu_1}\dots K_{\mu_{2n}}}{K^{2m}} = 
\sum_{k=0}^n 
\sum_{I\in \sigma_k}
\delta_{I^c}^0\, \delta_I^{i_1\dots i_{2k}} g_{i_1\dots i_{2k}}
\frac{\Gamma\left(\frac{d}{2}\right)}{2^n \Gamma\left(\frac{d}{2} + n\right)}
\sumint{K}\frac{\vec{K}^{2k} K_0^{2(n-k)}}{K^{2m}}\,,
\end{align}
where $I\in\sigma_k$ considers all possible subsets of $2k$ indices in the set $\boldsymbol{\mu}=\{\mu_1,\cdots,\mu_{2n}\}$ and $I^c=\boldsymbol{\mu} \setminus I$ is the complement of $I$. Moreover, we used the abbreviated notation $\delta^0_{a_1 \ldots a_m}\equiv\delta_{a_1}^0\dots \delta_{a_m}^0$ and $\delta^{b_1\ldots b_m}_{a_1 \ldots a_m}\equiv\delta_{a_1}^{b_1}\dots \delta_{a_m}^{b_m}$, and $g_{i_1\dots i_{2k}}$ denotes the fully symmetric tensor of rank $2k$. Finally, $K_0$ and $\vec{K}$ correspond, respectively, to the temporal and spatial parts of the loop momentum $K$. Applied to the simpler case in which $\boldsymbol{\mu}=\{\nu,\rho\}$, the expression above yields
\begin{align}
\sumint{K} \frac{K_\nu K_\rho}{K^{2m}} = \frac{1}{d}\delta^i_\nu \delta^j_\rho\, g_{ij}\,\sumint{K}\frac{\vec{K}^2}{K^{2m}} + \delta^0_\nu \delta^0_\rho\,\sumint{K}\frac{K_0^2}{K^{2m}}\,,
\end{align}
which is used in~\eqref{eq:TraceExpansion} for the toy-model example.

The remaining scalar sum-integrals are known (see, e.g.~\cite{Brauner:2016fla}). When the loop momentum is bosonic, which we denote by $[K]$ in the sum-integral, we have
\begin{align}\label{eq:BSumInt}
I^{\alpha\beta\gamma}_{\mathrm{B}}=\sumint{[K]}\frac{K^{2\beta}_0 \vec{K}^{2\gamma}}{K^{2\alpha}}= &\big(\overline{\Lambda}^2 C\big)^\epsilon\frac{2\Lambda_T}{(4\pi)^{d/2+1}}\left(\frac{\Lambda_T}{2}\right)^{d-2\alpha+2\beta+2\gamma}\notag\\
&\times\frac{\Gamma(\gamma+d/2)\Gamma(\alpha-\gamma-d/2)}{\Gamma(\alpha)\Gamma(d/2)}\,\zeta(2\alpha -2\beta-2\gamma -d)\,,
\end{align}
where $\zeta$ denotes the Riemann zeta function, $\overline{\Lambda}$ is the renormalization scale in the $\overline{\mathrm{MS}}$-scheme, and $C$ is a constant defined as
\begin{align}
C\equiv\left\{
\begin{array}{cc}
\frac{e^{-\gamma_E}}{4\pi}     &  \qquad d+2\epsilon\;\mathrm{odd}\\
\frac{e^{\gamma_E}}{(4\pi)^3}  &  \qquad d+2\epsilon\;\mathrm{even} 
\end{array}
\right.\,,
\end{align}
with $\gamma_E$ being Euler's constant. The sum-integrals for fermionic loop momenta, which we denote by $\{K\}$ in the sum-integral, are related to the bosonic ones as follows: 
\begin{align}\label{eq:FSumInt}
I^{\alpha\beta\gamma}_{\mathrm{F}}=\sumint{\{K\}}\frac{K^{2\beta}_0 \vec{K}^{2\gamma}}{K^{2\alpha}}=(2^{2\alpha-2\beta-2\gamma-d}-1)\, I^{\alpha\beta\gamma}_{\mathrm{B}}\,.
\end{align}

%%%%%%%%%%%%%%%%%%%%%%%%%%%%%%%%%%%%%%%%%%%%%%%%%%%%%%%%%%%%%%%%%%%%%%%%%%%%%%%%%%%%%%%%%%%%%%
\sectionlike{References}
\vspace{-10pt}
\bibliography{References} 	
%%%%%%%%%%%%%%%%%%%%%%%%%%%%%%%%%%%%%%%%%%%%%%%%%%%%%%%%%%%%%%%%%%%%%%%%%%%%%%%%%%%%%%%%%%%%%%

\end{document}